
\def\lsim{\mathrel{\scriptstyle{\buildrel < \over \sim}}}
\magnification 1200
\baselineskip=17pt


\centerline{\bf WIGNER CRYSTAL STATE FOR THE EDGE ELECTRONS }
\bigskip
\centerline{\bf IN THE QUANTUM HALL EFFECT AT FILLING $\nu = 2$}
\vskip 50pt
\centerline{J. P. Rodriguez,{\footnote {$^*$}
{Permanent address: Dept. of Physics
and Astronomy, California State University,
Los Angeles, CA 90032, USA.}}
M.J. Franco{\footnote {$^{\dag}$} {Present  address:
Meta4 Human Technology,
Centro Europa Empresarial - Edf. Roma,
 28230 Las Rozas, Madrid, Spain.}}
and L. Brey}
\medskip
\centerline{\it Instituto de Ciencia de Materiales de Madrid,
Consejo Superior de }
\centerline{\it Investigaciones Cient\'{\i}ficas,
Cantoblanco, 28049 Madrid, Spain}
\vskip 30pt
\centerline  {\bf  Abstract}
\vskip 8pt\noindent
The electronic excitations at the edges of a Hall bar 
not much wider than a few magnetic lengths
are studied theoretically
at filling $\nu = 2$.
Both mean-field theory and Luttinger liquid theory techniques
are employed for the case of   a null Zeeman energy splitting. 
The first calculation yields a stable spin-density wave state
along the bar, while the second one predicts dominant
Wigner-crystal correlations along the edges of the 
bar.  We propose an antiferromagnetic
Wigner-crystal groundstate for the edge electrons that
reconciles the two results.  A net Zeeman splitting
is found to produce canting of the antiferromagnetic order.

\bigskip
\noindent
PACS Indices: 73.40.Hm, 72.15.Nj
\vfill\eject
 
\centerline{\bf I. Introduction}
\bigskip
It was conjectured long ago by Wigner that a zero-temperature 
gas of electrons 
should transit into
a crystalline state as the electron density is lowered
to  sufficiently dilute levels.$^1$
The conjecture is based on the observation that the electronic
wavefunctions are localized 
when kinetic energy is absent from
the quantum mechanical Hamiltonian.
This  corresponds to the former dilute limit.$^2$
More recently, Schulz studied the electron gas confined to one dimension
through a bosonization analysis.$^3$  He found that the zero-temperature
groundstate is best described by a Wigner crystal state 
at {\it all} densities.
The complete  absence of a Luttinger liquid state,$^4$ which is the paradigm
for one-dimensional (1D) Fermi systems that experience
short-range interactions,
is consistent with the 
incomplete screening that is characteristic of the
electron gas confined to a 1D geometry.

Wigner crystals have also been proposed as groundstates of 
quantum Hall effect systems at sufficiently low 
two-dimensional (2D) electron densities.$^{5}$
Recently, two of the authors have shown that the auto-correlations
for $2k_F$ charge-density wave (CDW) order along the edges of a
thin quantum Hall
bar at electronic filling $\nu =1$
decay {\it slower} than any power law of the separation.$^{6}$
The groundstate is fully spin polarized in such case.
The  same dependence was obtained by Schulz for the Wigner
crystal ($4k_F$) auto-correlations in the case of the 
 electron gas confined to one dimension  in zero field.$^3$  It stands to 
reason then that the groundstate of the edge-electrons
for a thin quantum Hall bar
at filling $\nu =1$ is  effectively a spin-polarized 1D electron gas
with dominant Wigner-crystal correlations.

In this paper, we shall study the nature of the groundstate for the
edge electrons of a thin  quantum Hall bar at filling $\nu =2$,$^7$ in which
case both the spin up and the spin down lowest Landau levels are
completely filled.
Our main  conclusion is that such edge electrons
are best described by a 1D Wigner crystal state with
quasi long-range antiferromagnetic order for Hall bars of
widths not much larger than a few magnetic lengths. 
Evidence in support
of this claim has two sources.  First, we find that
a spin-density wave (SDW) state for the edge-electrons is
favored over a CDW one
in the absence of
Zeeman energy splitting at the mean-field level.
(A CDW groundstate for the edge-electrons was obtained
within the mean-field approximation
for the case of filling $\nu = 1$ in ref. 6.)  
The amplitude of the SDW is exponentially small for
Hall bars  of  widths much larger than a few magnetic lengths,
however.$^{8}$
This  calculation is presented in the next section.
Second, we obtain dominant Wigner-crystal
$4k_F$ correlations that decay {\it slower} than any power
law in the absence of Zeeman energy splitting 
   from a standard bosonization analysis of the Luttinger model
for the edge electrons.$^{6-8}$ 
Similar results were obtained  by Schulz for the
unpolarized  1D electron gas.$^3$
This is the subject of section III of the paper.
The two results just cited for the groundstate of
the edge-electrons  of
a quantum Hall bar  at filling $\nu =2$
appear  at first sight to be in conflict with each other.
To bring them into accord,    we then propose in section IV
that the groundstate for the edge electrons
is a 1D antiferromagnetic Mott insulator.$^2$
The previous restriction to zero Zeeman
energy splitting is relaxed at this stage, where the remaining spin-1/2
degrees of freedom are described by an effective 
antiferromagnetic chain in external 
magnetic field.  This yields a canted  1D Neel state
for the  ground-state spin configuration.
We find that the spin auto-correlations
of this model compare favorably with the SDW autocorrelations
obtained from the bosonization analysis of the Luttinger
model at null Zeeman splitting (see section III).
 This lends theoretical support to  our   proposal.

\bigskip
\bigskip
\centerline{\bf II. Mean-field Theory: CDW versus SDW}
\bigskip
As is well known, the quantum Hall effect is a   physical
realization of the two-dimensional (2D) electron liquid in strong magnetic
field.$^9$   The one-electron states are confined to the lowest
Landau level in such case.  Consider henceforth the open geometry
particular to a Hall bar of length $L_y$ and of width $W$.
We shall also assume periodic boundary conditions along both the
$x$ (width) and the $y$ (length) directions, with each  of
the respective  periods, $L_x$ and $L_y$, being
much greater than $W$.
 The quantum mechanical description of this system
 is therefore given by
the Hamiltonian $H = H_0 + H_1$, with kinetic energy
$$H_0 = \sum_{x,s} \varepsilon_{x,s} 
c_{x,s}^{\dag} c_{x,s} \eqno (1)$$
and with Coulomb interaction energy
$$H_1 = (2 L_x L_y)^{-1} \sum_{s_1, s_2}\sum_{x_1, x_2, \vec q}
V(\vec q) e^{iq_x (x_1 - x_2 - q_y l_0^2)} e^{-q^2 l_0^2/2}
c_{x_1,s_1}^{\dag}c_{x_2,s_2}^{\dag}
c_{x_2+q_y l_0^2,s_2}c_{x_1-q_y l_0^2,s_1}
\eqno (2)$$
expressed in second quantized form.$^{6}$
In particular, the operator $c_{x,s}^{\dag}$ creates a spin
$s = \uparrow,\downarrow$ electron in the lowest Landau level at
transverse  position $x$ within the Hall bar.  We
take the Landau gauge, which means that the longitudinal
momentum of this eigenstate  along the bar 
is equal to  $k_y = x/l_0^2$, where $l_0 = (\hbar c/eB)^{1/2}$
is the magnetic length.
The corresponding one-electron eigenenergy has the form
$$\varepsilon_{x,s} = \varepsilon_0(x) + s  E_Z/2,
\qquad (s = +, -)\eqno (3)$$
where $E_Z =  g_0 \mu_B H$ is the Zeeman energy splitting
of the lowest Landau level.
(The equivalent quantum numbers $x$ and $k_y$ will be freely
interchanged throughout the paper.)
The dispersion, $\varepsilon_0(x)$, of the Landau levels 
is dictated by the nature of the edge potentials (see Fig. 1 and
ref. 10).
Last, $V(q) = 2\pi e^2/\epsilon q$
above  represents the Fourier transform of
the repulsive Coulomb interaction, where
$\epsilon$ denotes the dielectric constant.
Depending on the  2D  electron density, $n$,
the lowest Landau level can be (fractionally) filled by a factor
of $\nu = 2\pi  l_0^2 n$.
The $\nu = 2$ state, for example, consists of completely filled lowest Landau
levels for both of the  spin-$1\over 2$ quantum numbers.
Below, we present evidence for the existence of
dominant SDW correlations along the  edges of the Hall bar
at this filling through a mean-field calculation.

Two of the authors have found theoretical evidence for a CDW instability
of the edge electrons in  the quantum Hall state at $\nu =1$,$^6$
in which case the lowest Landau level corresponding to
spin up is completely filled, while the spin down
level is empty.  We shall now determine the nature of
such Peierls instabilities in the case where both of the
lowest Landau levels are filled ($\nu =2$).
In order to simplify the analysis that follows, we shall first assume 
a null Zeeman energy splitting, $E_Z = 0$.  
This constraint will be relaxed later  in section IV.
Consider now  the CDW/SDW trial wavefunction
$$|DW_n\rangle =  \Pi_{k,s}
[u_k^{\prime} + s^n v_k^{\prime} c_{k+g(k),s}^{\dag}c_{k-g(k),s}]
\cdot
\Pi_{|q|\le k_F,s}c_{q,s}^{\dag}|0\rangle,\eqno (4)$$
with coherence factors $u_k^{\prime}$ and $v_k^{\prime}$, and with index
$n = 0,1$.
Here, $g(k) = ({\rm sgn}\, k)\, k_F$ denotes one-half 
of the CDW/SDW nesting vector.  
Its magnitude coincides with  the Fermi wave-number, 
$k_F = {1\over 2} W/l_0^2$, 
with respect to the momentum quantum number of the one-electron
states parallel to the edges of the Hall bar.  
In the trivial case that $u_k^{\prime} = 1$ and $v_k^{\prime} = 0$, 
we then  have a Fermi sea of free electron states localized 
at transverse position  $x$ that extends out
to the (Fermi) edges 
of the Hall bar: i.e., $|x| <  {1\over 2} W$
(see Fig. 1).
In the more general case $v_k^{\prime}\neq 0$, however,
$|DW_0\rangle$ represents a   CDW state, with nesting vector
$2 k_F$, while $|DW_1\rangle$ represents the corresponding   SDW state with
staggered magnetization oriented parallel to the 
spin quantization axis $\hat z$.
Note that the absence of Zeeman energy splitting
implies that the Hamiltonian $H_0+H_1$ has $SU(2)$ spin rotation
symmetry,
which guarantees that 
SDW states formed along any other quantization axis are degenerate.  
(It will be shown in section IV that the presence of a 
nonzero Zeeman energy splitting
favors {\it transverse} SDW states.)
The corresponding oscillations 
in the electronic density and in the electronic  spin that characterize
these states are computed   at the end of Appendix A. 
The CDW state, for example,
shows     ${\rm cos}(2k_F y)$ oscillations of the 
electron density at longitudinal
position $y$ along the Hall bar, while the SDW state
instead shows such oscillations in the
spin density  of the edge electrons [see Eq. (A16)].
Such oscillations are found to be
exponentially small, however, for Hall bars
of widths that are much greater than a few  magnetic lengths [see Eq. (A17)
and ref. 8].

To proceed further, it is
useful to observe that the groundstate wavefunction
(4) can be re-expressed as the product,
$$|DW_n\rangle =  \Pi_{k,s} A_{k,s}^{\dag}|0\rangle,\eqno (5)$$
of quasiparticle excitations
$$A_{k,s}^{\dag} = u_k c_{k-k_F,s}^{\dag}
                         +s^n v_k c_{k+k_F,s}^{\dag},\eqno (6)$$
with the restriction to wave-numbers
$|k| < k_F$.  This restriction shall be implicit in all
subsequent sums and products over quantum numbers.
Above, we have the identities 
$u_k = u_k^{\prime}$ and $v_k = v_k^{\prime}$
for $k > 0$, and
$u_k = v_k^{\prime}$ and $v_k = u_k^{\prime}$
for $k < 0$.
The normalization condition
$\langle DW_n|DW_n\rangle = 1$ thus imposes  the standard
constraint
$$|u_k|^2 + |v_k|^2 = 1 \eqno (7)$$
on the coherence factors.
Notice  that the quasiparticle creation operators then also
satisfy the normal   fermionic commutation relations
$$\{A_{i}, A_{j}^{\dag}\} =
\delta_{i,j}\eqno (8)$$
and
$$\{A_{i}, A_{j}\} 
= 0 = \{A_{i}^{\dag}, A_{j}^{\dag}\},
\eqno (9)$$
where $i =(k,s)$ and $j=(k^{\prime},s^{\prime})$
enumerate the corresponding quantum numbers.
We can  now make use of the identity (5) in conjunction with
the above commutation relations to compute the groundstate
energy of the trial density wave state (4).
The kinetic energy
$E_0 = \langle DW_n|H_0| DW_n\rangle$
for the CDW/SDW state is thereby obtained   
and found to take the form
$$E_0 = 2\sum_x \varepsilon_0(x)  n_x. \eqno (10)$$
Here 
$$n_x = |u_{x+k_F}|^2 + |v_{x-k_F}|^2 \eqno (11)$$
is the mean quasiparticle occupation.

The calculation of the Coulomb energy 
$E_1 = \langle DW_n|H_1| DW_n\rangle$ requires more 
work, however.  First, it is useful to note that this expectation value
is equal to the sum, $E_1 = E_{\rm dir} + E_{\rm exc}$, of a direct
Coulomb energy
$$E_{\rm dir} = \sum_{s_1, s_2} \sum_{x_1, x_2, q} 
B(q, x_1 - x_2)
\langle c_{x_1,s_1}^{\dag} c_{x_1-q,s_1} \rangle
\langle c_{x_2,s_2}^{\dag}c_{x_2+q,s_2} \rangle 
\eqno (12)$$
and of an exchange Coulomb energy
$$E_{\rm exc} = -\sum_{s_1, s_2} \sum_{x_1, x_2, q} 
B(q, x_1 - x_2)
\langle c_{x_1,s_1}^{\dag} c_{x_2+q,s_2} \rangle
\langle c_{x_2,s_2}^{\dag}c_{x_1-q,s_1} \rangle.
\eqno (13)$$
Here, we define the effective potential
$$B(q,x) = (2 L_x L_y)^{-1} 
\sum_k V(k,q) e^{ik(x-q)} e^{-(k^2+q^2)/2}.\eqno (14)$$
Above, and hence-forth, we measure all lengths in units of the
magnetic length: $l_0\rightarrow 1$.
After assuming the form (5) for the CDW/SDW groundstate wavefunction,
it is demonstrated in the Appendix A that the
corresponding  particle-hole expectation
values  that appear above are given by
$$\langle c_{x,s}^{\dag} c_{x^{\prime},s^{\prime}} \rangle
= \delta_{s,s^{\prime}}
 \sum_{w,w^{\prime} =  \pm } (w)_{x,s}^* 
(w^{\prime})_{x^{\prime},s^{\prime}}
\delta_{x, x^{\prime} + k_F (w - w^{\prime})},\eqno (15)$$
with the  coherence factors renamed 
$(-)_{x,s} = u_{x+k_F}$ and $(+)_{x,s} = s^n v_{x-k_F}$ 
for convenience [see Appendix A, Eqs. (A1)-(A7)].  
The product
$e_{\rm dir} = 
\langle  c_{x_1,s_1}^{\dag} c_{x_1-q,s_1} \rangle
\cdot
\langle  c_{x_2,s_2}^{\dag}c_{x_2+q,s_2}  \rangle$
that appears in expression (12)  for the direct 
Coulomb energy thus takes the form
$$ e_{\rm dir} =
 \sum_{w,w^{\prime} =  \pm }  (w)_{x_1,s_1}^*
(w^{\prime})_{x_1 - q,s_1}
\delta_{q, k_F (w - w^{\prime})}
\cdot
 \sum_{z,z^{\prime} =  \pm } (z)_{x_2,s_2}^*
(z^{\prime})_{x_2 + q,s_2}
\delta_{q, k_F (z^{\prime} -z)}. \eqno (16)$$
On the other hand, the  product
$e_{\rm exc} = \langle  c_{x_1,s_1}^{\dag} c_{x_2+q,s_2} \rangle
\cdot
\langle c_{x_2,s_2}^{\dag}c_{x_1-q,s_1} \rangle$
that appears in expression (13) for the exchange Coulomb energy
can be obtained from the direct term (16)
through    the replacement $q\rightarrow -q + x_1 - x_2$ 
in the case of like spins, $s_1 = s_2$
[see Appendix A, Eq. (A12)].  
It is zero otherwise.  
After some manipulation [see Appendix A, Eqs. (A11) and (A13)],
we therefore obtain the final form  
$$E_1 = \sum_{x_1,x_2} D_{x_1,x_2} n_{x_1} n_{x_2}
+ \sum_{x_1,x_2} D^{\prime}_{x_1,x_2}
{\rm Re} (u_{x_1}^* v_{x_1} u_{x_2} v_{x_2}^*)
\eqno (17)$$
for the total Coulomb interaction energy, with direct and exchange
kernals
$$\eqalignno{
D_{x_1,x_2}  =  & 4 B(0,x_1 - x_2) - 2\, B(x_1 - x_2, x_1 - x_2), & (18)\cr
D_{x_1,x_2}^{\prime}  =  & \delta_{n,0} \cdot  4
\sum_{w=\pm} B(2 k_F w, x_1 - x_2 +2 k_F w)\cr
 & \qquad\qquad
 - 2 \sum_{w=\pm} B(x_1 - x_2, x_1 -x_2 + 2 k_F w), & (19)\cr}$$
respectively.  
Observe that the exchange kernal  (19) diverges logarithmically at the Fermi
surface, $x_1 = 0 = x_2$, with an
attractive sign.  The divergent nature of the Peierls instability
present in the 1D Fermi sea ($v_k^{\prime} = 0$)
 then implies that the edge-electrons are
unstable to CDW/SDW condensates (4).
Yet the difference in energy between the CDW ($n = 0$) and
SDW ($n=1$) states (4) is  then equal to
$$E_{\rm CDW} - E_{\rm SDW} =  4
\sum_{x_1,x_2}   \sum_{w=\pm} B(2 k_F w, x_1 - x_2 +2 k_F w)
\cdot
{\rm Re} (u_{x_1}^* v_{x_1} u_{x_2} v_{x_2}^*)
\eqno (20)$$
for a given set of coherence factors.
The sum
$$\sum_{w=\pm} B(2k_F w, x_1 - x_2 +2k_F w) = 2  B(2k_F, x_1 - x_2 +2 k_F)$$
is positive, however, which yields the inequality
$$E_{\rm SDW} < E_{\rm CDW}.\eqno (21)$$  
If we now take the coherence factors to be those that minimize
the CDW energy functional $E_{\rm CDW} [u_k,v_k]$,
then the above inequality necessarily implies that
${\rm min}\,E_{\rm SDW} < {\rm min}\,E_{\rm CDW}$.
We conclude, therefore,
that  the SDW state is more stable
than the  CDW states within the present mean-field
theory approximation.

\bigskip
\bigskip
\centerline{\bf III. Luttinger Liquid Model}
\bigskip
The previous meanfield result indicates that SDW correlations along
the edges of a thin Hall bar are likely to be important at
filling $\nu =2$ in the absence of Zeeman splitting,
$E_Z =0$.  Recent   work by two of the authors indicates
that the groundstate of the edge electrons in the
fully spin-polarized
$\nu = 1$ state  is a
CDW,$^6$ which is consistent with this result.
These authors found, however, that the long-range CDW autocorrelations
decay {\it slower} than any power law, contrary to
what occurs in the case of short-range interactions.$^4$  In addition,
Schulz showed that the 1D electron gas is better described by a
Wigner crystal state than by a CDW or SDW one.$^3$
Could a Wigner crystal
state then be the true 
electronic groundstate at the edges of the quantum Hall
bar in the absence of
Zeeman splitting ($E_Z = 0$) for filling $\nu = 2$ as well?
Indeed, we shall demonstrate below that the
low-energy long-wavelength effective theory for
 the edge-electrons in such
case corresponds to that of a modified 1D electron gas.

In order to answer the question just posed,
we need  to obtain more information about      the dominant 
long-wavelength
electronic correlations at the edges of the Hall bar for
 filling $\nu = 2$. 
To this end, we introduce the following components
of the Luther-Emery (LE) model Hamiltonian,$^{11}$
$H = H_{\parallel} + H_{\perp}$, that can describe the
low-energy  dynamics
of the edge electrons at zero temperature
in the absence of Zeeman splitting:

$$\eqalignno{
H_{\parallel} =  \sum_{s = \uparrow,\downarrow}\int dy\Bigl[
 v_F \Bigl(L_s^{\dag}  i\partial_y L_s
 - R_s^{\dag} i\partial_y & R_s\Bigr)
 + g_{2,\parallel} L_s^{\dag}R_s^{\dag} R_s L_s +\cr
&+ g_{4,\parallel} \Bigl(L_s^{\dag} L_s L_s^{\dag} L_s 
+  R_s^{\dag} R_s  R_s^{\dag} R_s\Bigr) \Bigr], &  (22)\cr}$$

$$\eqalignno{
H_{\perp} =   \int dy \Bigl[ 
g_{1,\perp}  
\Bigl(
L_{\uparrow}^{\dag} R_{\downarrow}^{\dag} L_{\downarrow} R_{\uparrow}
 + {\rm h.c.}\Bigr)   
& + g_{2,\perp}
\Bigl(L_{\uparrow}^{\dag} L_{\uparrow}R_{\downarrow}^{\dag} R_{\downarrow} +
R_{\uparrow}^{\dag} R_{\uparrow}L_{\downarrow}^{\dag} L_{\downarrow}\Bigr) 
\Bigr]
 + \cr
&
 + g_{4,\perp}
\Bigl(L_{\uparrow}^{\dag} L_{\uparrow} L_{\downarrow}^{\dag} L_{\downarrow} +
R_{\uparrow}^{\dag} R_{\uparrow}R_{\downarrow}^{\dag} R_{\downarrow}\Bigr). 
& (23)\cr}$$
Here,
$e^{ik_F y} R_s (y) = L_y^{-1/2}\sum_k e^{iky} a_s(k)$ and
$e^{-ik_F y} L_s (y) = L_y^{-1/2}\sum_k e^{iky} b_s(k)$
denote field operators for right and left
moving spin $s$ fermions along the respective edges of the 
Hall bar.  In particular, $a_s(k)$ and $b_s(k)$ denote the
respective annihilation operators for spin $s$ edge electrons of momentum
$k$  that move
along the right and the left edges of the Hall bar,
with a Fermi surface at $\pm k_F$.
Note that point splitting and normal ordering
of the interaction terms in $H_{\parallel}$ and in $H_{\perp}$
is implicit where necessary.$^{4}$
Spin-charge separation then yields
the factorization    
$H_{\parallel} + H_{\perp} = H_{\rho} + H_{\sigma}$
of the electronic excitations
along the  coupled edges of the Hall bar,
where
$$\eqalignno{
H_{\rho} =  
& \sum_{q > 0}\sum_{j = R,L} 2\pi v_{\rho} (q)
\rho_j(q)\rho_j(-q)  + 
\sum_q g_{\rho} (q)\rho_R(q)\rho_L(-q) 
 & (24)\cr
H_{\sigma} =  
& \sum_{q > 0}\sum_{j = R,L} 2\pi v_{\sigma} (q)
\sigma_j(q)\sigma_j(-q) + 
\sum_q g_{\sigma} (q) \sigma_R(q)\sigma_L(-q) 
 + H_{\perp,1} & (25)\cr}$$
are the respective commuting portions of the Hamiltonian.
Here, 
$$\rho_j (q) = 2^{-1/2} [\rho_j(q,\uparrow) + \rho_j(q,\downarrow)]
\quad {\rm and} \quad
\sigma_j (q) = 2^{-1/2} [\rho_j(q,\uparrow) - \rho_j(q,\downarrow)]$$
are the standard particle-hole operators for collective charge and
spin excitations,
with $\rho_R (q,s) = L_y^{-1/2} \sum_k a_s^{\dag}(q+k) a_s(k)$ and
$\rho_L (q,s) = L_y^{-1/2} \sum_k b_s^{\dag}(q+k) b_s(k)$,
while
$H_{\perp,1} = g_{1,\perp}\sum_s\int dy 
L_{s}^{\dag} R_{-s}^{\dag} L_{-s} R_{s}$ 
is the LE spin-backscattering process.
Also, the Fermi velocities and interaction strengths for each
component
are renormalized by the forward scattering
processes [see Eqs. (22) and (23)]  to
$$\eqalignno{
v_{\rho,\sigma}  = 
&  v_F + 
 (g_{4,\parallel}  \pm g_{4,\perp})/2\pi, & (26)\cr
g_{\rho,\sigma} = 
& g_{2,\parallel}  \pm g_{2,\perp},  & (27)\cr}$$
where the $+(-)$ signs above correspond to the $\rho (\sigma)$ label.
Last, it is important to remark that $SU(2)$ spin rotation
 invariance yields   the
identity
$$g_{2,\parallel}  -   g_{2,\perp}  = - g_{1,\perp}\eqno (28)$$
between $g_{\sigma} = g_{2,\parallel}  -   g_{2,\perp}$
and the strength of the spin-backscattering interaction,
in addition to the identity
$$g_{4,\parallel} = g_{4,\perp}\eqno (29)$$
between the forward scattering processes.

Let us now determine the 
interaction parameters  that are theoretically expected for the above
LE model in the case
of the quantum Hall bar with null Zeeman splitting.$^{7}$
Since the Coulomb interaction is spin independent, we have the  
identities
$$\eqalignno{
g_{2,\parallel} = & {1\over 2} V_{\rm inter} = g_{2,\perp}
 & (30)\cr
g_{4,\parallel} = & {1\over 2} V_{\rm intra} = g_{4,\perp}
 & (31)\cr}$$
between the above LE model interaction parameters and the Fourier
transforms 
$$\eqalignno{
 V_{\rm intra}(q) & = 
{2 e^2\over\epsilon} {\rm ln} {\gamma_0\over {q W}} & (32)\cr 
V_{\rm inter}(q) & =
{2 e^2\over\epsilon} K_0 (qW) & (33)\cr}$$
of the  respective intra-edge and the inter-edge Coulomb potentials.
Above, $\gamma_0$ is a constant of order unity and
$K_0 (x)$ is a modified Bessel function.
These assignments then yield a wave-number dependent Luttinger liquid
for the charge sector (compare with ref. 6), parametrized by
$$v_{\rho} (q) = v_F +  V_{\rm intra} (q) /2\pi \quad
{\rm and} \quad g_{\rho}(q) = V_{\rm inter} (q),\eqno (34)$$
and  a {\it free} theory,
$$v_{\sigma} = v_F\quad
{\rm and} \quad g_{\sigma} = 0 = g_{1,\perp},\eqno (35)$$
for the spin sector.   The latter necessarily implies
that the spin gap is {\it null}.
As a first probe of the nature of this system, we shall
compute the static one-particle correlation function at
asymptotically large separations via the standard abelian
bosonization technique.$^{12}$
Spin-charge separation yields the form
$\langle \Psi_s(0) \Psi_s^{\dag} (y) \rangle = G_{\rho}(y) G_{\sigma} (y)$
for this quantity,
where 
$$\Psi_s(y) = L_s (y) + R_s (y) \eqno (36)$$ 
is the field operator that coherently removes
a spin $s$ edge electron at longitudinal position $y$ along the Hall bar.
The bosonization technique yields the expressions
$G_{\rho}(y) \sim  {\rm exp} [- {\rm const.}
\times ({\rm ln}\, y)^{3/2}/({\rm ln}\, W)^{1/2}]$
and $G_{\sigma} (y) \sim  (\alpha_0/y)^{1/2}$ for
the asymptotic one-particle autocorrelators of the charge and spin sectors,
respectively.  Here, $\alpha_0$ denotes the natural ultra-violet
length scale of the edge electrons.
The non-algebraic dependence of
$G_{\rho}(y)$ is a result of the wave-number dependence
(34) of the Luttinger model parameters 
due to the Coulomb interaction (see refs. 3 and 6).
We therefore obtain the asymptotic result
$$\langle \Psi_s(0) \Psi_s^{\dag} (y)\rangle \sim
(\alpha_0/y)^{1/2} {\rm exp}   [- {\rm const.}
\times ({\rm ln}\, y)^{3/2}/({\rm ln}\, W)^{1/2}]\eqno (37)$$
for the static one-particle autocorrelator, which notably
decays {\it faster} than any power law.  This clearly
indicates  that the edge electrons of the quantum Hall bar
at $\nu =2$ are not in a Luttinger liquid state in the
absence of Zeeman splitting, $E_Z = 0$.
[It should be remarked that the both the spin and the charge sectors
of the LE model,  Eqs. (24) and (25), exhibit Lorentz invariance.  This implies
that dynamic autocorrelations are related to the corresponding static
ones by a suitable identification of the time difference
$t$ with the relative displacement $y$. (See refs.  7, 8 and 12.)]

To gain yet more information about the nature of the correlated
state along the edges of the  $\nu =2$ quantum Hall bar,
we shall now compute the CDW and SDW auto-correlation
functions.  
Again, the bosonization technique    shall be employed.
The free spin sector (35) implies that the former are
identical ($K_{\sigma} = 1$, see ref. 4).  It is therefore
sufficient to compute the autocorrelator for the $2k_F$  CDW
order parameter,
 $O_{2} = R_{\uparrow}^{\dag} L_{\uparrow}
+ R_{\downarrow}^{\dag} L_{\downarrow}$.
It has the form
$\langle O_{2}(0) O_{2}^{\dag}(y)\rangle  =
{\rm cos}(2 k_F y) C_{\rho}(y) C_{\sigma}(y)$,
where$^6$
$C_{\rho}(y)  \sim {\rm exp}[-{\rm const.}^{\prime}\times 
({\rm ln}\, y\cdot {\rm ln}\,W)^{1/2}]$
and$^4$
$C_{\sigma}(y) \sim  \alpha_0/y$
are the respective charge and spin contributions.
This yields the result
$$\langle O_{2}(0) O_{2}^{\dag}(y)\rangle \sim
{\rm cos}(2 k_F y) \cdot (\alpha_0/y) 
\cdot {\rm exp}[-{\rm const.}^{\prime}\times
({\rm ln}\, y\cdot {\rm ln}\,W)^{1/2}] \eqno (38)$$
for the CDW/SDW autocorrelator asymptotically,  
which predominantly decays  as $1/y$ (see also ref. 8).  
Finally, we compute the auto-correlation function for the
$4k_F$ Wigner-crystal order parameter,
$O_4 = R_{\uparrow}^{\dag} R_{\downarrow}^{\dag}
L_{\downarrow} L_{\uparrow}$.
Application of the bosonization technique yields the
asymptotic form
$$\langle O_4(0) O_4^{\dag}(y)\rangle 
\sim {\rm cos}(4 k_F y)\cdot  {\rm exp}[- 2 \, {\rm const.}^{\prime}\times
({\rm ln}\, y\cdot {\rm ln}\,W)^{1/2}],\eqno (39)$$
which decays {\it slower} than any power law.$^3$
Notice that  sub-algebraic factors are
common to both  the CDW/SDW and to  the $4 k_F$
autocorrelators, (38) and (39). 
We conclude that Wigner crystal
correlations are dominant along the edges of the
quantum Hall bar in the absence of Zeeman energy splitting
at filling $\nu = 2$.

\bigskip
\bigskip
\centerline{\bf IV. Antiferromagnetic Wigner Crystal Proposal}
\bigskip

The prior Luttinger model analysis for the dynamics of
the edge electrons along  a
quantum Hall bar at filling $\nu =2$
in the absence of Zeeman energy splitting
 indicates that the corresponding groundstate is
in fact a 1D Wigner crystal of periodicity $a = \pi/2 k_F.$   Although
SDW correlations are also found to be quasi long range, 
they are subdominant.
The meanfield analysis presented in section II found dominant
SDW order with respect to CDW order, however.
These  results seem to be
incompatible at first site.   In principle, the meanfield result
could simply be wrong.  We believe this not to be
the case, however, for the following reasons.  It can be easily
shown that a meanfield analysis$^{13}$
 of the Luther-Emery model, (22) and (23),
in terms of CDW, SDW, and Cooper pair order parameters 
for the edge electrons 
recovers the same  phase diagram that is  obtained from
a renormalization-group analysis    of the equivalent Coulomb-gas
ensemble.$^{4, 12}$  This indicates that the mean-field analysis
made in section II for the entire quantum Hall bar
is correct in so far as the phase stability is concerned.

We propose instead  the following solution to the  puzzle outlined above:      
The groundstate of the edge electrons along  a
quantum Hall bar at $\nu =2$ is
{\it effectively} described by
 a spin-1/2 antiferromagnetic chain  in external magnetic field,
  with lattice constant 
$$a = \pi l_0^2/W.\eqno (40)$$
Notice that Eq. (40) indicates that
 the Mott localization effect$^2$ increases as
the width $W$ of the strip decreases.
The dominant Wigner crystal
correlations obtained in the
previous section indicate that the {\it effective} 1D spin-1/2 fermions
represented by the field operator $\Psi_s(y)$ [see Eq. (36)]
are localized at longitudinal points $y_i = i\cdot a$ along the Hall bar.
The remaining {\it effective} spin degree of freedom
$$\vec S_i = 
{1\over 2} \sum_{s,s^{\prime}} \Psi_s^{\dag}(y_i)\vec\sigma_{s,s^{\prime}}
\Psi_{s^{\prime}}(y_i)\eqno (41)$$
is a coherent measure of the spin along both edges of the Hall bar.  It
is then governed by the Hamiltonian for the
antiferromagnetic Heisenberg chain ($J > 0$) in external magnetic field
$\vec h =  E_Z \hat z$:$^{14}$
$$H_{\sigma} = J\sum_i \vec S_i\cdot \vec S_{i+1} - \vec h\cdot 
\sum_i \vec S_i.\eqno (42)$$
The spin-wave excitations of the chain also disperse linearly,$^{15}$
with  a group velocity $v_{\sigma}^{\prime} =\pi J a/2 \hbar$
in zero field.
The latter is obtained from comparison with the exact Bethe ansatz
solution    for the spin-1/2 Heisenberg chain.$^{16}$
Notice that the absence of a spin gap is consistent with
the previous Luttinger model analysis.
Indeed, comparison with Eq. (35) in turn yields the formula
$$J = (2/\pi)^2 \hbar k_F  v_F \eqno (43)$$
for the effective exchange-coupling constant in zero field.

The above spin-1/2 antiferromagnetic chain (42) is well understood
theoretically.  
The groundstate is a 
Neel state (plus zero-point spin-wave fluctuations) that is
canted in the direction of the external field.  
Indeed, the  application  of the Jordan-Wigner transformation to this spin
chain yields an equivalent system
of interacting 1D spinless fermions.$^{17}$ 
A subsequent Luttinger model analysis  
then yields asymptotic  transverse and
longitudinal static spin correlations
$$\eqalignno{
\langle S_i^{+} S_{i+m}^{-}\rangle \sim & (-1)^m/m & (44)\cr
\langle\delta  S_i^z \delta S_{i+m}^z \rangle \sim & 
[{\rm cos}(2 k_F a \cdot m)] / m, & (45)\cr}$$
at low fields
that are consistent with such a canted antiferromagnetic state
(see Appendix B).
Above, $\delta S_z$ refers to the deviation of the magnetization
with respect to the mean spin polarization
$$\langle S^z_i\rangle = \chi_{\perp} h_z  \eqno (46)$$
along the direction $\hat z$ of the applied field,
while $k_F\cong  \pi/2a$ 
denotes the field-dependent Fermi wavenumber of the
1D spinless fermions.
Here
$\chi_{\perp} =  1/ \pi^2 J$ is the transverse magnetic susceptibility
known exactly at zero field.$^{16}$
In particular, the incommensurate SDW (45) represents the 
effect of canting,
whereas the transverse spin correlations (44) describe the antiferromagnetic
component of the configuration.
Observe that the spin auto-correlators, (44) and (45), are quite
similar to the SDW auto-correlator (38) obtain previously from
the Luttinger model analysis for the edge-electron dynamics
in the absence of Zeeman splitting.
The only difference between the two results is the absence of
an additional sub-algebraic factor in the spin auto-correlators for
the antiferromagnetic chain.  This is likely to be connected with
the fact that the analog of the 
$4k_F$ autocorrelator (39)  in the
spin-chain model (42) is simply the  density-density correlation function,
which is identically equal to unity, and which therefore  shows no
such sub-algebraic dependence.
Last, the canted Neel state saturates to a fully polarized
state at  a  critical field $h_c\sim J$.  Comparison of
the edge band structure depicted in Fig. 1 with
Eq. (43) indicates, however, that the system has passed into
a $\nu = 1$ quantum Hall state by then.  
The later nevertheless possesses a ferromagnetic
groundstate spin configuration and 
Wigner-crystal-type correlations
along the edges.$^6$  This coincides    precisely with the nature of the
saturated canted antiferromagnetic state at $h_c$.

Given the above Wigner-crystal proposal, we can now go back
and make contact with the previous mean-field calculation presented in
section II.  In the limit of small Zeeman
 field, $\vec h\rightarrow 0$,
the change in energy of a SDW state in comparison to its
zero-field value takes the bilinear form
$$\delta E_{\rm SDW}/{\cal N} = 
-{1\over 2}
( \chi_{\parallel} h_{\parallel}^2 + \chi_{\perp} h_{\perp}^2)
\eqno (47)$$
in terms of the parallel and perpendicular paramagnetic susceptibilities
and fields with respect to the staggered magnetization of the SDW.
Here ${\cal N}$ denotes the total number of conduction electrons in
the Hall bar.
The presence of a single-particle gap in the mean-field
excitation spectrum implies that the longitudinal susceptibility
is null, however, at zero temperature: $\chi_{\parallel} = 0$.  
Also, comparison with
the previous antiferromagnetic chain yields  a non-zero transverse
paramagnetic susceptibility, $\chi_{\perp} \sim J^{-1}$.
Eq. (47) then indicates that the lowest-energy SDW states
are those with the staggered magnetization oriented
{\it perpendicular} to the external field: e.g.
$$|SDW_x\rangle =  \Pi_{k,s}
[u_k^{\prime} +  v_k^{\prime} c_{k+g(k),s}^{\dag}c_{k-g(k),-s}]
\cdot
\Pi_{|q|\le k_F,s}c_{q,s}^{\dag}|0\rangle.\eqno (48)$$
Examination of Fig. 1 reveals that such mean-field SDW
states permit a unique definition  of the nesting vector, $2 k_F$.
The free Fermi sea thus experiences a Peierls instability
that mixes the spin quantum number following Fig. 1.
Notice, however, that a unique nesting vector {\it cannot}
be defined in either the case of the CDW or of  the longitudinal
SDW states, $|DW_0\rangle$ and $|DW_1\rangle = |SDW_z\rangle$, for
$E_Z > 0$.  This is undoubtedly the principal reason why
such states are not stable in general.
[We remind the reader that the longitudinal (4) and
transverse (48) SDW states are degenerate in the absence
of Zeeman energy splitting due to $SU(2)$ spin rotation
symmetry.]

\bigskip
\bigskip
\centerline{\bf V. Conclusions}
\bigskip
By combining the results obtained from a mean-field theory  and a
Luttinger liquid analysis of the edge-electrons of a thin quantum
Hall bar at filling $\nu = 2$, we propose  that the corresponding
groundstate is in a Wigner crystal state, with the remaining spin
degrees of freedom arranged into a canted antiferromagnetic configuration.
The amplitude of  both the respective    CDW and SDW is exponentially
small, however, for Hall bars of width much larger than a few magnetic
lengths [see Appendix A, Eqs. (A16) and (A17), and ref. 8].  
We remind the reader that two of the authors have shown that a
similar Wigner crystal/CDW state is expected theoretically
along the edges of a quantum Hall bar in the fully spin-polarized
state found at filling $\nu = 1$.$^6$

There are various experimental tests that could identify the above
Wigner-crystal canted antiferromagnetic state at $\nu = 2$.  First,
it is known that
any 1D Wigner crystal is pinned by a single ``backscattering''
impurity.$^{18,19}$  In the present context, such a defect
corresponds to a correlated pin that extends in the transverse
direction  from one edge of
the Hall bar to the other.  Nonlinear conduction along
the length of the Hall bar
is then expected below a threshold voltage, while ohmic conduction
is expected above this threshold.  The former corresponds to
a pinned Wigner
crystal state while the latter corresponds to
a sliding Wigner crystal state.
Second, a good probe for the
 predicted SDW order is the spin-lattice relaxation
rate, $1/T_1$, in  nuclear magnetic resonance.$^8$  This
quantity is directly proportional to the transverse spin-correlator (44),
which means that it is sensitive to 
the canted antiferromagnetic order predicted
here theoretically.

J.P.R. thanks E. Rezayi for discussions.  He 
also acknowledges partial support from the 
National Science Foundation 
through grant \# DMR-9322427, and partial
support     from the Spanish Ministry for
Education and Culture.
L.B. was supported in part by 
the Spanish Ministry for
Education and Culture under contract \# PB 960085.

\vfill\eject
\centerline{\bf Appendix A: Evaluation of 
$\langle c_{x^{\prime},s^{\prime}}^{\dag} c_{x,s}\rangle$, etc.}
\bigskip
Eqs. (12) and (13) indicate that
the essential object in  the evaluation of the Coulomb energy (2) with
respect to the CDW/SDW groundstate [Eqs. (4) and (5)] is the 
particle-hole expectation value
$$\langle DW_n|c_{x^{\prime},s^{\prime}}^{\dag} c_{x,s}
|DW_n\rangle = \langle 0 |\Pi_{i^{\prime}}
 A_{i^{\prime}}\cdot
c_{x^{\prime},s^{\prime}}^{\dag} c_{x,s}\cdot
\Pi_{i} A_{i}^{\dag}|0\rangle. \eqno (A1)$$
The latter can be computed by progressively commuting the annihilation
operator $c_{x,s}$ to the right,
and by progressively commuting
the creation operator $c_{x^{\prime},s^{\prime}}^{\dag}$
to the left above.  This process results in
the identity
$$c_{x,s}| DW_n\rangle =
\sum_i (-1)^{i+1} \{c_{x,s}, A_{i}^{\dag}\}
\Pi_{j\neq i} A_j^{\dag}|0\rangle \eqno (A2)$$
for the correlated  hole excitation.  Evidently, it is a coherent
sum of all possible hole excitations weighted
by the appropriate commutator
$$\{c_{x,s}, A_{i}^{\dag}\} =
\delta_{s,s_i}(u_{k_i}\delta_{x, k_i-k_F} 
+ s_i^n v_{k_i}\delta_{x, k_i+k_F}).\eqno (A3)$$
Use of the identity 
$\langle 0|A_i A_j^{\dag}|0\rangle = \delta_{i,j}$
for the quasiparticle excitations
then yields the expression
$$\langle DW_n|c_{x^{\prime},s^{\prime}}^{\dag} c_{x,s}
|DW_n\rangle
= \sum_i \{ A_i, c_{x^{\prime},s^{\prime}}^{\dag}\}
\{c_{x,s}, A_{i}^{\dag}\} \eqno (A4)$$
for the expectation value (A1).
Finally, substitution of (A3) above yields
the desired result
$$\langle DW_n|c_{x^{\prime},s^{\prime}}^{\dag} c_{x,s}
|DW_n\rangle
=  \delta_{s,s^{\prime}}
 \sum_{w,w^{\prime} =  \pm } (w)_{x^{\prime},s^{\prime}}^*
(w^{\prime})_{x,s}
\delta_{x^{\prime}, x + k_F (w - w^{\prime})} \eqno (A5)$$
for the expectation value in terms of the coherence factors, which have been
renamed as
$$\eqalignno{
(-)_{x,s} &  = u_{x+k_F} & (A6)\cr
(+)_{x,s} &  = s^n v_{x-k_F} & (A7)\cr}$$
for convenience.

We shall next evaluate the products
$$\eqalignno{
e_{\rm dir} = & 
\langle DW_n| c_{x_1,s_1}^{\dag} c_{x_1-q,s_1}|DW_n \rangle
\cdot
\langle DW_n| c_{x_2,s_2}^{\dag}c_{x_2+q,s_2} |DW_n \rangle
& (A8)\cr
e_{\rm exc} = &
\langle DW_n| c_{x_1,s_1}^{\dag} c_{x_2+q,s_2}| DW_n \rangle
\cdot
\langle DW_n|c_{x_2,s_2}^{\dag}c_{x_1-q,s_1} | DW_n \rangle
& (A9)\cr}$$
that appear in expressions (12) and (13) for the direct and exchange
contributions to the Coulomb interaction energy.
Substitution of the prior result (A5) into (A8), for example, yields
the expression
$$ e_{\rm dir} = 
 \sum_{w,w^{\prime} =  \pm }  (w)_{x_1,s_1}^*
(w^{\prime})_{x_1 - q,s_1}
\delta_{q, k_F (w - w^{\prime})}
\cdot
 \sum_{z,z^{\prime} =  \pm } (z)_{x_2,s_2}^*
(z^{\prime})_{x_2 + q,s_2}
\delta_{q, k_F (z^{\prime} -z)}  \eqno (A10)$$
for the direct term.  After some manipulation, one obtains the
final result
$$\eqalignno{
e_{\rm dir} = \delta_{q,0 } & \sum_{w = \pm} 
|(w)_{x_1,s_1}|^2  
\sum_{w^{\prime} = \pm}
 |(w^{\prime})_{x_2,s_2}|^2 +
\cr 
& + \sum_{w = \pm}  \delta_{q, 2k_F w}
(w)_{x_1,s_1}^* (\bar w)_{x_1-2 k_F w,s_1}
(\bar w)_{x_2,s_2}^* (w)_{x_2+2 k_F w,s_2} & (A11)\cr}$$
for the product (A8), where we define the notation $\bar w = -w$.
In the case of the exchange term (A9), on the other hand, substitution
of (A5) yields the expression
$$\eqalignno{
 e_{\rm exc} = \delta_{s_1,s_2}
 \sum_{w,w^{\prime} =  \pm }  (w)_{x_1,s_1}^*
& (w^{\prime})_{x_2 + q,s_2}
\delta_{x_1 - x_2 - q, k_F (w - w^{\prime})}
\times\cr
& \times
 \sum_{z,z^{\prime} =  \pm } (z)_{x_2,s_2}^*
(z^{\prime})_{x_1 - q,s_1}
\delta_{x_1 -x_2 - q, k_F (z^{\prime} -z)}. &     (A12)}$$
Again, after some manipulation, we arrive at the result
$$\eqalignno{
e_{\rm exc} = \delta_{s_1,s_2}
\Biggl(\delta_{q,x_1 - x_2 } & \sum_{w = \pm} 
|(w)_{x_1,s_1}|^2 
\sum_{w^{\prime} = \pm} |(w^{\prime})_{x_2,s_2}|^2 +
\cr
& + \sum_{w = \pm} \delta_{q, x_1 - x_2 - 2 k_F w}
 (w)_{x_1,s_1}^* (\bar w)_{x_1-2 k_F w,s_1}
(\bar w)_{x_2,s_2}^* (w)_{x_2+2 k_F w,s_2}\Biggr).\cr
& & (A13)\cr}$$
Notice that $e_{\rm exc}$ can be obtained from $e_{\rm dir}$
by making the replacement
$q\rightarrow -q + x_1 - x_2$ in the case of
like spins, $s_1 = s_2$ [see Eqs. (A8) and (A9)].

Last, we shall compute the expectation value of
the number/spin density operator
$$S^{(\mu)}(\vec r) = \sum_{s,s^{\prime}} \sum_{k,k^{\prime}}
a_{s,s^{\prime}}^{(\mu)}
\langle k|\vec r\rangle \langle\vec r|k^{\prime}\rangle
c_{k,s}^{\dag} c_{k^{\prime},s^{\prime}}\eqno (A14)$$
at a point $\vec r = (x,y)$ in the Hall bar.  Here
$a^{(1)}$, $a^{(2)}$ and $a^{(3)}$ are the Pauli
matrices corresponding to the spin density, while
$a_{s,s^{\prime}}^{(0)} = \delta_{s,s^{\prime}}$ 
corresponds to the number density.  
Also, the wavefunction 
$\langle k|\vec r\rangle = L_y^{-1/2}e^{iky}\phi_k(x)$
represents an electron with momentum $k$ along the right edge
of the Hall bar in the Landau gauge, with the guiding center
factor 
$\phi_k(x) = \pi^{-1/4} l_0^{-1/2} {\rm exp}[-(x - k l_0^2)^2/2 l_0^2]$.
Use of the result (A5)
for the particle-hole amplitude yields the expression 
$$\langle S^{(\mu)}(\vec r)\rangle  = \sum_{s} \sum_{k,k^{\prime}}
a_{s,s}^{(\mu)}
\langle k|\vec r\rangle \langle\vec r|k^{\prime}\rangle
 \sum_{w,w^{\prime} =  \pm } (w)_{k,s}^*
(w^{\prime})_{k^{\prime},s}
\delta_{k, k^{\prime} + k_F (w - w^{\prime})},\eqno (A15)$$
for the expectation value of the number/spin density (A14).
After some manipulation, we obtain the final form
$$\eqalignno{
 \langle S^{(\mu)}(\vec r)\rangle  = & L_y^{-1}\sum_{k,s}
a_{s,s}^{(\mu)}
\Bigl[|\phi_{k-k_F}(x)|^2 |u_k|^2 + |\phi_{k+k_F}(x)|^2 |v_k|^2\Bigr] +\cr
& + 2\,{\rm cos}(2 k_F y + \phi) L_y^{-1}\sum_{k,s}
[a_{s,s}^{(\mu)}\cdot s^n] 
\phi_{k+k_F}(x) \phi_{k-k_F}(x) |u_k^* v_k|, & (A16)\cr}$$
where $\phi$ is the relative phase in between the
coherence factors at $k = 0$.
Since $\sum_{s} a_{s,s}^{(0)}\cdot s^n = 2$
and  $\sum_{s} a_{s,s}^{(3)}\cdot s^n = 0$
for the CDW state, $|DW_0\rangle$,
and vice versa for the SDW state,  $|DW_1\rangle$,
we conclude that the CDW state shows 
${\rm cos}(2 k_F y + \phi)$ oscillations in the electronic density along the
Hall bar, while the SDW state shows such oscillations in the electronic
spin density along the $z$-axis 
[see the contributions to the second term in Eq. (A16) near $k=0$].
The amplitude of such oscillations is limited by the product
$$\phi_{k+k_F}(x) \phi_{k-k_F}(x) =\pi^{-1/2} l_0^{-1}
 e^{-[(x-k l_0^2)/l_0]^2 - (k_F l_0)^2}\eqno (A17)$$
of the guiding center wavefunctions at the opposing edges,
which is appreciable for
thin Hall bars, $W \lsim 2  l_0$.$^8$

\bigskip\bigskip
\centerline{\bf Appendix B:  
$S = 1/2$ Antiferromagnetic Chain in External Field}
\bigskip
We shall now  reproduce the well-known
bosonization analysis$^{14}$ of the $s=1/2$ antiferromagnetic chain (42)
for  the purpose of  showing
that the groundstate in external field is in a canted Neel
configuration.  It is useful to consider first the more general case
with anisotropy described by the Hamiltonian
$$H_{\sigma} = \sum_i\Bigl[{1\over 2} 
J_{xy}  (S_i^{+} S_{i+1}^{-} + S_{i+1}^{+} S_i^{-}) + 
J_z S_i^z S_{i+1}^z - h_z S_i^z\Bigr]\eqno (B1)$$
for the  $s=1/2$ XXZ chain in external magnetic field $h_z$ aligned
parallel to the anisotropy axis.  Application of the Jordan-Wigner
transformation$^{17}$
$$S_i^{-} = f_i e^{i\pi N_i}\quad {\rm and}\quad 
S_i^{+} = f_i^{\dag} e^{-i\pi N_i}\eqno (B2)$$
of the lowering and raising operators
to spinless fermions, $f_i$, plus the string
$N_i = \sum_{j < i} f_j^{\dag} f_j$ yields the equivalent
fermion representation
$$H_{\sigma} = \sum_i\Bigl[{1\over 2} 
J_{xy}  (f_i^{\dag} f_{i+1} + {\rm h.c.}) +
J_z n_i n_{i+1} - h_z n_i\Bigr]\eqno (B3)$$
of the spin Hamiltonian (B1).  Above, we have used the identity
$$S_i^z = n_i - {1\over 2}\eqno (B4)$$
in between the spin operator along the anisotropy axis and
the occupation number $n_i  =  f_i^{\dag} f_i$ for the
spinless fermions.  This system is a compressible Luttinger liquid for
$J_{xy} \geq J_z$, which translates into a paramagnetic response
$$\langle S_i^z\rangle = \chi_{\perp} h_z.\eqno (B5)$$  
Comparison of the
latter  with
Eq. (B4) then indicates that the system of spinless fermions (B3)
is half filled at zero field, and off half filling at
$h_z\neq 0$.  Following Luther and Peschel,$^{14}$ the correlation functions
can be easily calculation via the bosonization technique, where
one obtains the results
$$\eqalignno{
\langle S_i^{-} S_{i+x/a}^{+}\rangle & \sim
(-1)^{x/a} \Bigl({\alpha_0\over x}\Bigr)^{(2K)^{-1}}, & (B6)\cr
\langle \delta S_i^z \delta S_{i+x/a}^z\rangle & \sim
{\rm cos}(2k_F x) \Bigl({\alpha_0\over x}\Bigr)^{2K} & (B7)\cr}$$
in the asymptotic limit $x\rightarrow\infty$.
Above, $k_F$ denotes the field-dependent 
Fermi wavenumber, while $K$ denotes the
characteristic Luttinger liquid parameter.
Also, $\alpha_0^{-1}\sim a^{-1}$ is the momentum cutoff of the
underlying Luttinger model, while
$\delta S_i^z = S_i^z - \langle S_i^z\rangle $.
Notice that the string factors (B2)
remove any dependence of the oscillatory factor
in the transverse 
spin autocorrelator (B6) with $k_F$.

In the case of
the Heisenberg chain, $J_{xy} = J_z$, the spin correlations are isotropic
at zero magnetic field.  This requires that  
$2K = 1 = (2K)^{-1}$ in such case.  We therefore obtain the spin
autocorrelations
$$\eqalignno{
\langle S_i^{-} S_{i+m}^{+}\rangle & \sim
(-1)^{m}/m , & (B8)\cr
\langle \delta S_i^z \delta S_{i+m}^z\rangle & \sim
[{\rm cos}(2k_F a \cdot m)]/m & (B9)\cr}$$
in the low field limit,
where $k_F \cong \pi/2a$.  Study of Eqs. (B5), (B8) and (B9) then indicates
that the  groundstate of the $s=1/2$ antiferromagnetic Heisenberg
chain  in the presence of a weak external magnetic field is in
a canted Neel configuration.

\vfill\eject
\centerline{\bf References}
\vskip 16 pt
\item {1.}  E. Wigner, Phys. Rev. {\bf 46}, 1002 (1934).

\item {2.}  N.F. Mott, {\it Metal-Insulator Transitions}
(Taylor and Francis, London, 1990).


\item {3.} H.J. Schulz, Phys. Rev. Lett. {\bf 71}, 1864 (1993).

\item {4.} J. Voit, Rep. Prog. Phys. {\bf 58}, 977 (1995).

\item {5.} H.E. Fertig in  {\it Perspectives in Quantum Hall Effects},
edited by S. Das Sarma and A. Pinczuk (Wiley 1997).

\item {6.} M. Franco and L. Brey, Phys. Rev. Lett. {\bf 77}, 1358 (1996).

\item {7.} H.C. Lee and S.-R. Eric Yang, Phys. Rev. B {\bf 56}, 15529
(1997).

\item {8.} H.C. Lee and S.-R. Eric Yang, Phys. Rev. B
 {\bf 57}, R4249 (1998).

\item {9.} {\it The Quantum Hall Effect}, edited by R.E. Prange
and S.M. Girvin, (Springer-Verlag, 1990), 2nd ed..


\item {10.}  B.I. Halperin, Phys. Rev. B {\bf 25}, 2185 (1982);
 X.G. Wen, Int. J. Mod. Phys. B {\bf 8}, 457 (1994).

\item {11.} A. Luther and V.J. Emery, Phys. Rev. Lett. {\bf 33}, 589 (1974).

\item {12.} V.J. Emery, in {\it Highly Conducting One-dimensional Solids},
ed. by J.T. Devreese, R.P. Evrard and V.E. van Doren
(Plenum Press, New York, 1979).

\item {13.} J.P. Rodriguez, Europhys. Lett. {\bf 39}, 195 (1997);
 Europhys. Lett. {\bf 47}, 745 (E) (1999);
Phys. Rev. B {\bf 58}, 944 (1998).

\item {14.} A. Luther and I. Peschel, Phys. Rev. B {\bf 12}, 3908 (1975).

\item {15.} F.D.M. Haldane, Phys. Rev. Lett. {\bf 45}, 1358 (1980).

\item {16.} R.J. Baxter, Ann. Phys. (N.Y.) {\bf 70}, 193 (1972).

\item {17.} E. Fradkin, {\it Field Theories of Condensed Matter Systems}
(Addison-Wesley, Redwood City, 1991), chap. 4.

\item {18.} C.L. Kane and M.P.A. Fisher, Phys. Rev. B {\bf 46},
15233 (1992).

\item {19.} L.I. Glazman, I.M. Ruzin and B.I. Shklovskii,
Phys. Rev. B {\bf 45}, 8454 (1992).

\vfill\eject
\centerline{\bf Figure Caption}
\vskip 20pt
\item {Fig. 1.}  Shown is the schematic band structure of the edge
electrons along the quantum Hall bar at filling $\nu = 2$.  The 
horizontal arrows
represent the unique nesting vector, $2 k_F$, for the transverse
SDW states [see Eq. (48), for example].


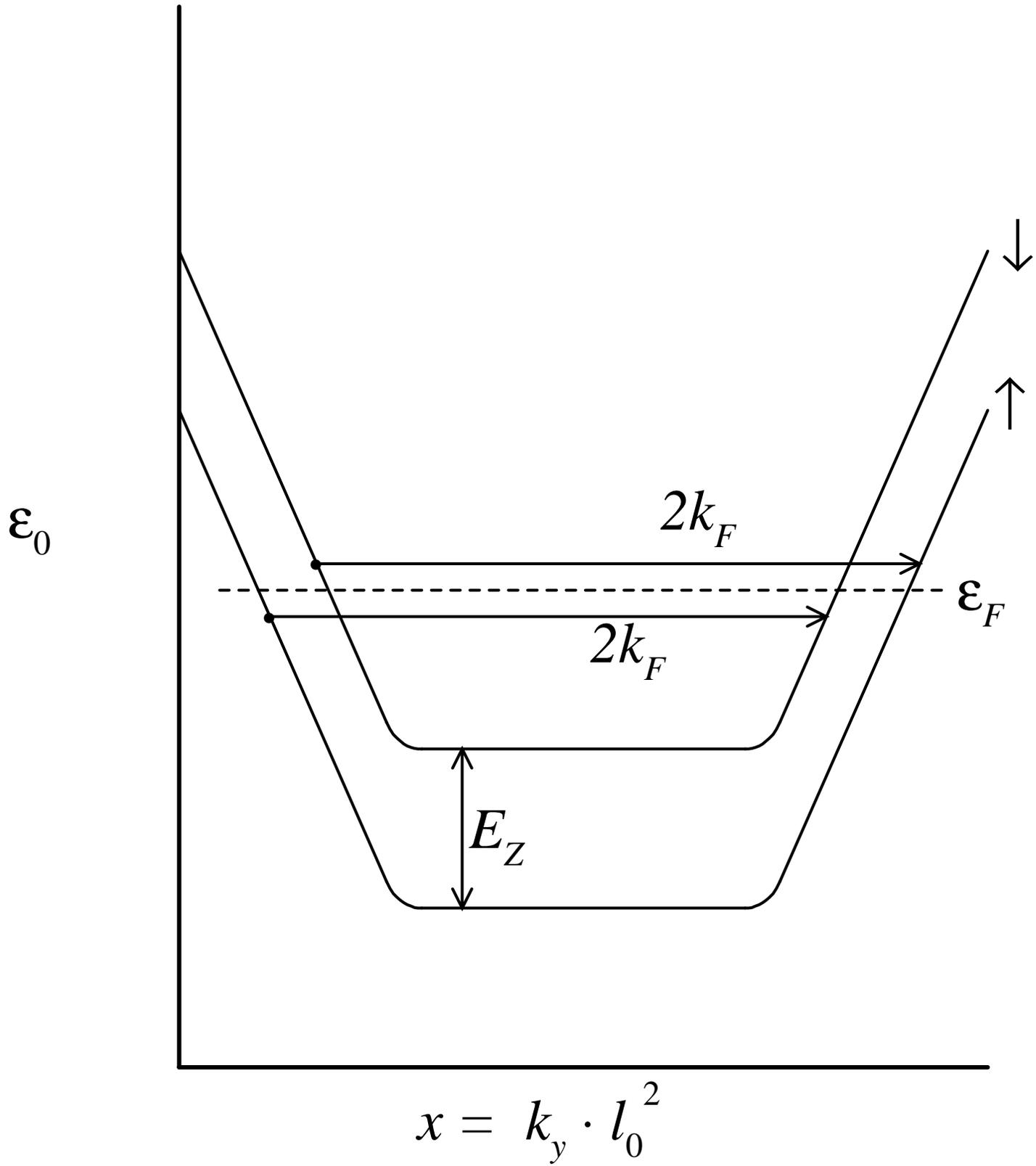
\end